\documentclass[11pt]{article}
\usepackage{graphicx}
\usepackage{amssymb}
\usepackage{amsmath}
\usepackage[square,numbers,sort&compress]{natbib}
\newcommand{\bm}[1]{\mbox{\boldmath${#1}$}}

\def\nv{N_3} \def\nvmu{N^\mu_3}
\def\na{N_4}
\def\namu{N^\mu_4}
\def\dela{\Delta_4^i}
\def\delamu{\Delta^{\mu i}_4}

\def\inner#1#2{{\bm #1}\cdot {\bm #2}}

\def\Iprot#1#2{P^{#1#2}_{3/2}}
\def\Iproo#1#2{P^{#1#2}_{1/2}}
\def\Sprot#1#2{\Gamma^{#1#2}_{3/2}}
\def\Sproo#1#2{\Gamma^{#1#2}_{1/2}}
\newcommand{\del}{\partial}
\def\half{\frac{1}{2}}
\begin{document}
\title{Chiral Properties of Baryon Interpolating Fields}
\author{Keitaro Nagata~\footnote{Address after the 20th
May 2007, Department of Physics, Chung-Yuan Christian University,
Chung-Li 320, Taiwan} , Atsushi Hosaka,\\
{\it Research Center for Nuclear Physics,  Osaka University}\\
{\it Ibaraki 567-0047, Japan,}\\
and \\
V. Dmitra\v sinovi\' c,\\
{\it Institut J. Stefan, Dept. of Theoretical Physics F-1,}\\
{\it  Jamova 39, 1000 Ljubljana, Slovenia;} \\
{\it Permanent address: Vin\v ca Institute of Nuclear Sciences, lab 010} \\
{\it P.O.Box 522, 11001 Beograd, Serbia. }}
\date{}
\maketitle
%
\begin{center}
{\bf Abstract}
\end{center}
We study the chiral transformation properties of all possible
local (non-derivative) interpolating field operators for baryons
consisting of three quarks with two flavors, assuming good isospin
symmetry. We derive and use the relations/identities among the
baryon operators with identical quantum numbers that follow from
the combined colour, Dirac and isospin Fierz transformations.
These relations reduce the number of independent baryon operators
with any given spin and isospin. The Fierz identities also
effectively restrict allowed baryon chiral multiplets. 
It turns out that the chiral multiplets of the baryons 
are equivalent to their Lorentz representation.
For the two independent nucleon operators the only permissible chiral
multiplet is the fundamental one $(\frac12,0)\oplus(0,\frac12)$.
For the $\Delta$, admissible Lorentz representations are $(1,\frac12)\oplus
(\frac12,1)$ and $(\frac32,0)\oplus(0,\frac32)$. In the case of
the $(1,\frac12)\oplus (\frac12,1)$ chiral multiplet the
$I(J)=\frac32(\frac32)$ $\Delta$ field has one
$I(J)=\frac12(\frac32)$ chiral partner; otherwise it has none. We
also consider the Abelian ($U_A(1)$) chiral transformation
properties of fields and show that each baryon comes in two
varieties: 1) with Abelian axial charge +3; and 2) with Abelian
axial charge -1. In case of the nucleon these are the two Ioffe's
fields; in case of the $\Delta$, the $(1,\frac12)\oplus
(\frac12,1)$ multiplet has Abelian axial charge -1 and the
$(\frac32,0)\oplus(0,\frac32)$ multiplet has Abelian axial charge
+3.

\section{Introduction}

Chiral symmetry is one of the global symmetries of QCD and plays a
key role in hadron physics. In the real world, chiral symmetry is
spontaneously broken by the QCD ground state (``vacuum"), where a
chiral condensate acquires a finite value. It is believed that the
chiral symmetry is restored at sufficiently high
density/temperature. Properties of hadrons would be modified under
such circumstances, and the chiral symmetry, as an almost exact
symmetry, would become important in the classification of hadron
spectra: in particular, one would observe a degeneracy of chiral
partners.

We call chiral partners a set of hadrons that share a certain
representation $(I_R,\; I_L)$ of the chiral group $SU(2)_R\times
SU(2)_L$, where $I_{R,L}$ label the representations of the right-,
and left- isospin groups $SU(2)_{R,L}$.
In the Wigner-Weyl phase, all the hadrons that fall into one
chiral multiplet, have their masses and axial-couplings determined
by the chiral symmetry.
Clear identification of chiral partners would
greatly facilitate finding evidence for the chiral restoration, if
there is any.

The best known example is the $\sigma$ as the chiral partner of
the $\vec{\pi}$ meson, the quartet of fields $(\sigma,\;
\vec{\pi})$ forming the chiral multiplet $(\frac12,\frac12)$.
Another example is the set of $(\vec{\rho},\;\vec{a}_1)$ mesons,
which are described by the interpolating fields $\vec{\rho} \sim
\bar{q} \gamma^\mu \vec{\tau} q,\; \vec{a}_1 \sim \bar{q}
\gamma^\mu\gamma_5 \vec{\tau} q$ that form the chiral multiplet
$(1,0) \oplus (0,1)$.

In contrast to the case of mesons, chiral partners of baryons are
not well established. Even the mere existence of chiral partner(s)
of the the ground state baryon, the nucleon, is still an open
question.

The main difficulty in finding the chiral partners of the nucleon
is the lack of knowledge of chiral multiplets of baryons, because
the observed baryon resonances do not manifestly belong to
irreducible chiral representations in the spontaneously broken
phase of chiral symmetry. Rather, a physical hadron is expected to
be described by a linear combination of several chiral
representations. At the level of the hadrons, one must
assume/guess their chiral representations.

On the other hand, when one describes hadrons starting from the
underlying quark substructure, the chiral multiplet
(representation) for the hadrons can be determined uniquely, as
shown in the case of the mesons, for instance $\sigma,\;\vec{\pi}$ and
$\vec{\rho},\;\vec{a_1}$.

There is an infinite number of possibilities for baryon
interpolating fields as, for instance one need not have only three
quarks, but also five, seven or more. In addition, one may include
(covariant) derivatives, or gluon fields into the baryon
operators. Many lattice and QCD sum rule studies suggest, however,
that the local, i.e. without derivatives, baryon operators
consisting of three quarks successfully describe properties of the
ground state baryons.

The construction of local operators for baryons was first studied
by Ioffe~\cite{Ioffe81}, Chung et al.~\cite{Chung:1981cc} and by
Espriu et al.~\cite{Espriu:1983hu}. These authors showed that the
Fierz transformations provide relations/identities among baryon
operators with identical spin and isospin, and thus reduce the
number of independent operators, e.g. they famously reduce the
number of independent nucleon fields from five to
two~\cite{Ioffe81,Espriu:1983hu}, and the number of independent
$\Delta$ fields from two to one (when one neglects fields
belonging to the Lorentz group representation $D(\frac32,0)$).

The first systematic study of chiral multiplets of baryon
operators was attempted by Cohen and Ji~\cite{CohenJi}. They did
not explicitly discuss the Fierz identities among various fields,
however, albeit they were aware of their existence. It is, in
fact, necessary to consider the Fierz transformations, because
they effectively restrict the allowed chiral multiplets of baryon
operators.

Baryon interpolating fields are constructed in such a way that
they belong to the same (irreducible) representations of the
Lorentz and of the isospin $SU(2)$ group. It is not {\it a priori}
obvious, however, that they also belong to the irreducible
representations of the chiral group $SU(2)_R\times SU(2)_L$. We
shall show that the Fierz identities/relations among baryon
operators lead precisely to the (non-)vanishing of those linear
combinations that form various chiral multiplets/representations.
This should not be surprising as the Fierz identities form an
implementation of the Pauli principle, and different permutation
symmetry classes form distinct multiplets of composite particles.
Hence it is necessary to take into account the Fierz identities
among baryon operators.

The standard isospin formalism greatly facilitates derivation of
the Fierz identities and chiral transformations of baryon
operators, due to the fact that both the quarks and the nucleons
belong to the iso-doublet representation. The composite Fierz
identities (i.e. in both the Dirac and isospin space) and the
chiral transformations of baryons are straightforwardly derived
using the iso-doublet representation. We also consider the Abelian
($U_A(1)$) chiral transformations of baryons and show that most
baryons come in two varieties: 1) with axial charge +3; and 2)
with axial charge -1.

This paper is organized as follows. In section~\ref{sec:baryon},
we first define all possible quark bi-linear fields and summarize
their chiral transformations. With quark bi-linear fields, all
possible baryon operators can be systematically defined. We
classify the baryon operators according to the representations of
the Lorentz and the isospin groups. Then we derive the Fierz
identities among the baryon operators for each representation of
the Lorentz and isospin group. In section~\ref{sec:chiral_baryon},
we derive the Abelian and non-Abelian chiral transformations of the
baryon operators as functions of the quarks' chiral transformation
parameters, using the iso-doublet representation. All possible
chiral multiplets for the baryon operators are enumerated by
taking into account the Fierz identities. The final section is a
summary and an outlook to the future.

\section{Baryon Operators}
\label{sec:baryon}

Local interpolating operators for baryons consisting of three
quarks can be generally described as
\begin{eqnarray}
B(x)\sim \epsilon_{abc} \left(q^T_a(x) \Gamma_1 q_b(x)\right) \Gamma_2 q_c(x),
\label{eq:bgeneral}
\end{eqnarray}
where $q(x)=(u(x),\;d(x))^T$ is an iso-doublet quark field at
location $x$, the superscript $T$ represents the transpose and the
indices $a,\;b$ and $c$ represent the color. Hence the
antisymmetric tensor in color space $\epsilon_{abc}$, ensures the
baryons' being color singlets. From now on, we shall omit the
color indices and space-time coordinates. $\Gamma_{1,2}$ describe
the tensor product of Dirac and isospin matrices. With a suitable
choice of $\Gamma_{1,2}$, the baryon operators are defined so that
they form an irreducible representation of the Lorentz and isospin
groups, as we shall show in this section.

Note that we employ the iso-doublet form for the quark field $q$,
although the explicit expressions in terms of up and down quarks
are usually employed in lattice QCD and QCD sum rule studies. We
shall see that the iso-doublet formulation leads to a simple
classification of baryons into isospin multiplets and to a
straightforward derivation of Fierz identities and chiral
transformations of baryon operators.

We begin with bi-linears of two quarks in Eq.~(\ref{eq:bgeneral}).
It is convenient to introduce a ``tilde-transposed" quark field
$\tilde{q}$ as follows
\begin{eqnarray}
\tilde{q}=q^T C\gamma_5 (i\tau_2),
\end{eqnarray}
where $C = i\gamma_2\gamma_0$ is the Dirac field charge
conjugation operator, $\tau_2$ is the second isospin Pauli matrix,
whose elements form the antisymmetric tensor in isodoublet space.
Taking into account the Pauli principle, there are five
non-vanishing possibilities for $\Gamma_1$ (otherwise there would
have been twice as many):
\begin{subequations}
\begin{eqnarray}
D_1&=&\tilde{q}q,\\
D_2&=&\tilde{q}\gamma^5 q,\\
D_3^\mu&=&\tilde{q}\gamma^\mu q,\\
D_4^{\mu i}&=&\tilde{q}\gamma^\mu\gamma^5\tau^i q,\\
D_5^{\mu\nu i}&=&\tilde{q}\sigma^{\mu\nu}\tau^i  q.
\end{eqnarray}
\label{eq:diquark5}
\end{subequations}
These quark bi-linear fields, $D_1$, $D_2$, $D_3^\mu$, $D_4^{\mu
i}$ and $D_5^{\mu\nu i}$, are referred to as the scalar,
pseudo-scalar, vector, axial-vector and tensor diquarks, by their
Lorentz transformation properties~\footnote{Throughout the present
paper, Latin indices $i,\;j$ etc. run over the isospin space
$1,\;2,\;3$, and Greek indices $\alpha,\beta$ etc. run over the
Lorentz space $0,\;1,\;2,\;3$.}.

Firstly we investigate their Abelian chiral ($U(1)_A$)
transformation properties, which is given by $U(1)_A$
transformation of the quark,
\begin{eqnarray}
q\to \exp(i\gamma_5 a) q,\\
\tilde{q} \to \tilde{q} \exp(i\gamma_5 a),
\end{eqnarray}
where $a$ is an infinitesimal parameter for $U(1)_A$.
The scalar and pseudo-scalar diquarks transform as
\begin{eqnarray}
\delta_5 D_1= 2ia D_2,\\
\delta_5 D_2= 2ia D_1,
\end{eqnarray}
the vector and axial-vector diquarks,
\begin{eqnarray}
\delta_5 D_{3,4}= 0,
\end{eqnarray}
the tensor diquark,
\begin{eqnarray}
\delta_5 D_5^{\mu \nu i}=2ia D_6^{\mu \nu i},
\end{eqnarray}
where ${D}^{\mu\nu i}_6=\tilde{q}\sigma^{\mu\nu}\gamma_5 \tau^i q$
is a dual-tensor diquark. Note that baryon operators constructed
from the dual-tensor diquark can be expressed as functions of the
tensor diquark by using the identity
$\sigma^{\mu\nu}\gamma_5=-\frac{i}{2}\epsilon^{\mu\nu\alpha\beta}
\sigma_{\alpha\beta}$.

We consider the chiral ($SU(2)_A$) transformation
\begin{subequations}
\begin{eqnarray}
{q} &\to& \exp (i \gamma_{5} \inner{\tau}{a}){q} ,
\label{eq:Qtrfch}\\
{\tilde q} &\to& {\tilde q}\exp (-i \inner{\tau}{a}\gamma_{5}),
\end{eqnarray}%
\label{eq:Qtrf}%
\end{subequations}%
where $\vec{a}$ is the  triplet of $SU(2)_A$ group parameters.
It is straightforward to obtain the chiral transformations of the diquarks:
for the scalar and pseudo-scalar diquarks,
\begin{eqnarray}
\delta_5^{\vec{a}} D =0,\; (D=D_1,D_2),\label{eq:chirald1}
\label{eq:chirald2}
\end{eqnarray}%
for the vector and axial-vector diquarks,
\begin{eqnarray}
\delta_5^{\vec{a}} D^\mu_3 =2i a^i D^{\mu i}_4,
\label{eq:chirald3}\\
\delta_5^{\vec{a}} D^{\mu i}_4 =2 i a^i D^\mu_3,
\label{eq:chirald4}
\end{eqnarray}%
for the tensor diquark,
\begin{eqnarray}
\delta_5^{\vec{a}} D_5^{\mu\nu i}=-2 \epsilon^{ijk} a^j D_6^{\mu\nu k}.
\label{eq:chirald5}
\end{eqnarray}
The scalar and pseudo-scalar diquarks $D_{1,2}$ are chiral scalars
(invariants) $(0,\; 0)$.  The vector and axial-vector diquarks
$D_{3}^\mu,\; D_4^{\mu i}$ together belong to the chiral multiplet
$(\frac12,\;\frac12)$; therefore they are chiral partners, similar
to the $(\sigma,\;\vec{\pi})$ case. The tensor diquark transforms
into the dual-tensor diquark, and they together form the chiral
multiplet $(1,0)\oplus (0,1)$.

Now we proceed to the baryon operators:
There we consider the Pauli principle in two steps. The first step
is the Pauli principle applied to the first and second quarks,
i.e. to the diquarks, as already performed and discussed above.
Second, additional constraint comes from the permutation of the
second and the third quark, which corresponds to the Fierz
transformation.
Note that the Fierz transformation connects only baryon operators
belonging to the same Lorentz and isospin group multiplets.
Therefore, we may classify the baryon operators according to their
Lorentz and isospin representations following Chung et
al~\cite{Chung:1981cc}. It has been known that such baryon
operators may couple either to the even or to the odd parity
states. In the following discussions all the baryon operators will
be defined as having even parity. We note, however, that the
baryon operators belonging to the same chiral multiplet may have
either parity.

Firstly, we consider the simplest case $D(\frac12,0)_{I=\frac12}$,
where $D(\frac12,0)$ denotes the representation of the Lorentz
group and $I=\frac12$ denotes the isospin. There are five
differently-looking operators,
\begin{eqnarray}
N_1&=&(\tilde{q}q)q,\\
N_2&=&(\tilde{q}\gamma_5 q) \gamma_5 q,\\
N_3&=&(\tilde{q}\gamma_\mu q)\gamma^\mu q,\\
N_4&=&(\tilde{q}\gamma_\mu\gamma_5\tau^i q) \gamma^\mu\gamma_5\tau^i q,\\
N_5&=&(\tilde{q} \sigma_{\mu\nu}\tau^i q) \sigma^{\mu\nu} \tau^i q,
\end{eqnarray}
where the subscripts $1,2,\cdots, 5$ describe their diquark
components of Eq. (3). As a consequence of the Fierz
transformations, see Appendix \ref{app}, we obtain three
identities:
\begin{eqnarray}
3 N_3&=&-N_4,
\label{eq:fierza1} \\
N_4&=&3(N_2-N_1),
\label{eq:fierzb1} \\
N_5&=&6(N_1+N_2). \label{eq:fierzn1}
\end{eqnarray}
Therefore, only two among the five operators are independent under
the Pauli principle, as noted by Ioffe \cite{Ioffe81} and by
Espriu et al. \cite{Espriu:1983hu}.

Next we consider $D(\frac12,0)_{I=\frac32}$. Baryon operators with
$I=\frac32$ must contain either the axial-vector or the tensor
diquark, so there are only two possibilities,
\begin{eqnarray}
\Delta^i_4&=&(\tilde{q}\gamma_\mu \gamma_5\tau^j q)
\gamma^\mu\gamma_5\Iprot{i}{j} q,\\
\Delta_5^i&=&(\tilde{q} \sigma_{\mu\nu}\tau^j q)\sigma^{\mu\nu} \Iprot{i}{j}q.
\end{eqnarray}
Here $\Iprot{i}{j}$ is the isospin-projection operator for
$I=\frac32$, which is defined, together with an isospin-projection
operator $\Iproo{i}{j}$ for $I=\frac12$, as
\begin{eqnarray}
\Iprot{i}{j}=\delta^{ij}-\frac13 \tau^i\tau^j,\;
\Iproo{i}{j}=\frac13 \tau^i\tau^j.
\end{eqnarray}
The $I=\frac32$ projection operator satisfies $\tau^i
P^{ij}_{\frac32}=0$, which ensures $\tau^i\Delta^i_{4,5}=0$. Using
the appropriate Fierz transformations, see Appendix \ref{app}, we
find that both these operators vanish,
\begin{eqnarray}
\Delta_4^i=0,
\label{eq:fierzc1} \\
\Delta_5^i=0. \label{eq:fierzd1}
\end{eqnarray}
The vanishing of these operators implies that the nucleon
operators belong only to the fundamental chiral representation, as
we shall show in the next section.
Moreover, the vanishing of the $I=\frac32(J^{P}=\frac12^{+})$
spatially symmetric states is in accord with the non-relativistic
result.

For $D(1,\frac12)_{I=\frac12}$, baryon operators may contain the
vector and the axial-vector, or the tensor diquark. Hence there
are three operators
\begin{eqnarray}
N_3^\mu&=&(\tilde{q}\gamma_\nu q) \Sprot{\mu}{\nu} q,\\
N_4^\mu&=&(\tilde{q}\gamma_\nu\gamma_5 \tau^i q) \Sprot{\mu}{\nu}\gamma_5
\tau^i q,\\
N_5^{\mu}&=&i(\tilde{q} \sigma_{\alpha\beta}\tau^i q)\Sprot{\mu}{\alpha}
\gamma^\beta\tau^i q,
\label{eq:baryonN5mu}
\end{eqnarray}
where the imaginary identity $i$ is introduced into the definition
Eq.~(\ref{eq:baryonN5mu}) in order to maintain the reality of all
coefficients in the Fierz identities, see Appendix \ref{app}.
Similarly to the isospin projection operators, $\Sprot{\mu}{\nu}$
is the spin-projection operator for $J=\frac32$ states, which is
defined, together with the $J=\frac12$ projection operator
$\Sproo{\mu}{\nu}$, by
\begin{eqnarray}
\Sprot{\mu}{\nu}=g^{\mu\nu}-\frac14 \gamma^\mu\gamma^\nu,
\; \Sproo{\mu}{\nu}=\frac14 \gamma^\mu\gamma^\nu.
\end{eqnarray}
Owing to this projection operator, the $J=\frac32$ baryon
operators satisfy the Rarita-Schwinger condition $\gamma_\mu
N_{3,4}^\mu=0$. Here we comment that a vector-spinor
(Rarita-Schwinger) field must satisfy another condition, $\del_\mu
N^{\mu}=0$, that can be satisfied by employing another spin
projection operator that contains
derivatives~\cite{Benmerrouche:1989uc}, which we do not use here,
in compliance with the standard usage in lattice QCD and QCD sum
rules. The Fierz transformation provides two relations
\begin{eqnarray}
N_3^\mu &=& N_4^\mu,
\label{eq:fierznmu1}\\
N_5^\mu &=& - 2N_3^\mu. \label{eq:fierznmu2}
\end{eqnarray}
There is therefore one independent $J=\frac32(I=\frac12)$
operator.

For $D(1,\frac12)_{I=\frac32}$, there are two operators
\begin{eqnarray}
\Delta^{\mu i}_4&=&(\tilde{q}\gamma_\nu\gamma_5\tau^j q)
\Sprot{\mu}{\nu}\gamma_5\Iprot{i}{j} q,\\
\Delta^{\mu i}_5&=&i (\tilde{q} \sigma_{\alpha\beta}\tau^j q)
\Sprot{\mu}{\alpha}\gamma^\beta\Iprot{i}{j} q.
\end{eqnarray}
We obtain the Fierz identity
\begin{eqnarray}
\Delta^{\mu i}_4 &=& \Delta^{\mu i}_5 .
\label{eq:fierznmu3}
\end{eqnarray}
Therefore, there is only one independent $J=\frac32(I=\frac32)$,
$D(1,\frac12)_{I=\frac32}$ operator, in accord with Ioffe's claim
\cite{Ioffe81}. This is not the only possible
$J=\frac32,I=\frac32$ operator, however.

There is another $J=\frac32$ operator in the
$D(\frac32,0)_{I=\frac12}$ Lorentz representation, that is a
Lorentz tensor
\begin{eqnarray}
N^{\mu\nu}_{5}&=& (\tilde{q} \sigma_{\alpha\beta}\tau^i q)
\Gamma^{\mu\nu\alpha\beta} \tau^i  q\, ,
\end{eqnarray}
where $\Gamma^{\mu\nu\alpha\beta}$ is another $J=\frac32$
projection operator defined as
\begin{eqnarray}
\Gamma^{\mu\nu\alpha\beta}=\left(g^{\mu\alpha}g^{\nu\beta}
-\half g^{\nu\beta}\gamma^\mu\gamma^\alpha
+\half g^{\mu\beta}\gamma^\nu\gamma^\alpha
+\frac16 \sigma^{\mu\nu}\sigma^{\alpha\beta}\right),
\end{eqnarray}
which ensures that $N^{\mu\nu}_5 = - N^{\nu\mu}_5,\; \gamma_\mu
N^{\mu\nu}_5 = 0$. Using the Fierz transformation we obtain
$N_5^{\mu\nu} = 0$, i.e., this field vanishes due to the Pauli
principle.

Finally in the $D(\frac32,0)_{I=\frac32}$ Lorentz representation,
there is one $\Delta$ operator
\begin{eqnarray}
\Delta^{\mu\nu i}_{5}&=& (\tilde{q} \sigma_{\alpha\beta}\tau^j q)
\Gamma^{\mu\nu\alpha\beta} \Iprot{i}{j}  q.
\label{e:chung}
\end{eqnarray}
The Fierz transformation generates only the trivial relation
$\Delta_5^{\mu\nu i} = \Delta_5^{\mu\nu i}$, so this operator
survives the total anti-symmetrization and provides a rarely
considered alternative for the $\Delta(1232)$ interpolating field.

We have defined various local operators for
three-quark baryons by classifying them according to their Lorentz
and isospin group representations. We have explicitly shown that
the Fierz transformation connects the baryon operators with
identical Lorentz and isospin properties. In the next section, we
shall show that the Fierz identities restrict the Pauli-allowed
baryons and hence also restrict their chiral multiplets.
In this sense, the chiral $SU(2)_R \times SU(2)_L$ symmetry plays
a similar role in the relativistic treatment of baryons, to the
one of $SU_{FS}(6)$ symmetry in the nonrelativistic treatment.
Presumably, one can apply this form of the Pauli principle also to
other, non-local three-quark fields that correspond to the
orbitally excited states, but we leave that problem for another
occasion.

\section{Chiral Transformations}
\label{sec:chiral_baryon}
In this section, we investigate the chiral transformations of
three-quark baryon operators. The chiral mixing of baryon
operators is caused by their diquark components, so it is
convenient to classify the baryon operators according to their
diquark chiral multiplets: $D_1$ and $ D_2\in (0,\;0)$, $D_3^\mu$
and $D_4^{\mu i} \in (\frac12,\;\frac12)$ and $D_5^{\mu\nu i}\in
(1,\;0)+(0,\;1)$.

First, we consider the baryon operators $N_1$ and $N_2$ that are
constructed from the scalar and pseudo-scalar diquarks. As shown
in Eq.~(\ref{eq:chirald1}), the scalar and pseudo-scalar diquarks
are chiral scalars $(0,0)$, therefore $N_1$ and $N_2$ belong to
the fundamental representation, whose transformations are simply,
\begin{eqnarray}
\delta_5^{\vec{a}}N &=& i \gamma_5 \inner{\tau}{a} N,\;~~~ (N =
N_1,\;N_2).
\end{eqnarray}
Under the Abelian chiral transformation the rule is also linear, but more
complicated as it mixes the two nucleon fields:
\begin{eqnarray}
\delta_5 N_{1} &=& i a \gamma_5 (N_{1} + 2 N_{2})
\label{e:nuc+Atrf1} \\
\delta_5 N_{2} &=& i a \gamma_5 (N_{2} + 2 N_{1}).
\label{e:nuc+Atrf2}
\end{eqnarray}
In other words these two nucleon fields appear to form a two-dimensional
representation of the Abelian chiral symmetry $U_{L}(1) \times
U_{R}(1)$, or an $U_{A}(1)$ doublet. Of course, all irreducible
representations of any Abelian Lie group are one-dimensional, so
the ``chiral doublet" two-dimensional representation of $U_{A}(1)$
furnished by the fields $N_{1,2}$ and defined by Eqs.
(\ref{e:nuc+Atrf1}) and (\ref{e:nuc+Atrf2}) must be a reducible
one. We may perform the reduction by taking the symmetric and
antisymmetric linear combinations of two 
nucleon fields $N_{1,2}$:
\begin{eqnarray}
N_{n} &=&  (N_{1} + N_{2}) \\
N_{m} &=&  (N_{1} - N_{2}).
\label{e:def}
\end{eqnarray}
Then their Abelian chiral transformation properties are
\begin{eqnarray}
\delta_5 N_{n} &=& 3 i a \gamma_5 N_{n}
\label{e:nuc+AtrfA} \\
\delta_5 N_{m} &=& - i a \gamma_5 N_{m} , \label{e:nuc+AtrfB}
\end{eqnarray}
Note the factor 3 in front of the r.h.s. of Eq.
(\ref{e:nuc+AtrfA}), i.e., it is the ``triply naive" Abelian axial
baryon charge transformation law, as it should be for an object
consisting of three quarks, and the negative sign in front of the
r.h.s. of Eq. (\ref{e:nuc+AtrfB}), as it should for an Abelian
``mirror" nucleon\footnote{We refer to such fields that transform
with the positive sign on the r.h.s. as ``naive" or covariant, and
to those that transform with the negative sign as ``mirror" or
contra-variant.}.

With the vector and axial-vector diquarks, we can construct six
kinds of baryon operators, four of them are $I=\frac12$:
$\nv,\;\nvmu,\;\na$ and $\namu$, and two of them $I=\frac32$:
$\dela$ and $\delamu$. The $SU(2)_A$ transformation defined by
Eq.~(\ref{eq:Qtrfch}) can change the isospin of the baryon
operators, but it cannot change its spin $J$, as familiar from the
case of mesons e.g. $(\sigma\;,\vec{\pi})$. Therefore, we may
divide chiral transformations of these fields into two sets: one
is the set of $J=\frac12$  $(\nv,\;\na,\;\dela)$ fields and the
other is the set of $J=\frac32$  $(\nvmu,\; \namu,\;\delamu)$.

The $SU(2)_A$ transformations of the first set ($J=\frac12$) are
given by
\begin{subequations}
\begin{eqnarray}
\delta_5^{\vec{a}} \nv &=& - i \inner{a}{\tau}\gamma_5 \nv -
\frac23 i\inner{a}{\tau}
\gamma_5\na - 2i \gamma_5 \inner{a}{\Delta_4},\\
\delta_5^{\vec{a}} \na &=& - 2i \inner{a}{\tau}\gamma_5 \nv
+ \frac13 i\inner{a}{\tau}\gamma_5 \na - 2 i\gamma_5\inner{a}{\Delta_4},\\
\delta_5^{\vec{a}} \dela &=& -2i\gamma_5 a^j \Iprot{i}{j}\nv
-\frac23 i\gamma_5 a^j P^{ij}_{3/2} \na + \frac23 i \tau^i
\gamma_5 \inner{a}{\Delta_4} - i\inner{a}{\tau}\gamma_5 \dela.
\end{eqnarray}
\label{eq:trva1}
\end{subequations}%
As mentioned earlier, these baryon operators transform only one
$J=\frac12$ field into another, i.e., the $J=\frac12$ operators
close this algebra and there is no mixing of the $J=\frac12$ and
$J=\frac32$ operators.

We find that Eqs.~(\ref{eq:trva1}) can be reduced to irreducible
components by taking the antisymmetric linear combination of the
two nucleon fields:
\begin{eqnarray}
\delta_5^{\vec{a}} (\nv - \na) &=& i\inner{a}{\tau} \gamma_5 (\nv-\na),\\
\delta_5^{\vec{a}}(3\nv + \na) &=& -\frac53 i
\inner{a}{\tau}\gamma_5 (3\nv+\na)
- 8 i\gamma_5 \inner{a}{\Delta_A},\\
\delta_5^{\vec{a}} \dela &=& -\frac23i\gamma_5 a^j P^{ij}_{3/2}
(3\nv + \na) + \frac23 i \tau^i \gamma_5 \inner{a}{\Delta_A} -
i\inner{a}{\tau}\gamma_5 \dela,
\end{eqnarray}
i.e., $(\nv-\na)$ forms the fundamental representation
$(\frac12,\;0) \oplus (0,\;\frac12)$, whereas $(3\nv+\na)$
together with $\dela$ form the higher dimensional representation
$(1,\;\frac12)\oplus (\frac12,\;1)$: $3\nv+\na$ and $\dela$ appear
to be chiral partners. The Fierz identities
Eqs.~(\ref{eq:fierza1}), (\ref{eq:fierzb1}), and
(\ref{eq:fierzc1}), however, forbid exactly the appearance of the
$(1,\;\frac12)\oplus (\frac12,\;1)$ baryon fields/chiral partners,
as the said identities make precisely these (linear combinations
of) fields vanish identically.

Hence the only possible chiral multiplet containing the nucleon
field is the fundamental one, i.e.,
\begin{eqnarray}
\delta_5^{\vec{a}} N &=& i\inner{a}{\tau}\gamma_5 N,~~~ (N = N_3,\;N_4),
\end{eqnarray}
while the Abelian chiral transformation is given by
\begin{eqnarray}
\delta_5 N &=& - i \gamma_5 a N,~~~ (N=N_3,\;N_4),
\end{eqnarray}
hence both of these nucleon fields have the single-mirror property
under the Abelian chiral transformation.

Similarly to the first set, at first sight the $J=\frac32$ fields
may be in both the $(\frac12,0)\oplus(0,\frac12)$ and in the
$(1,\frac12)\oplus(\frac12,1)$ multiplet. In this case, however,
the Fierz identities Eq.~(\ref{eq:fierznmu1}) forbid the
fundamental representation, leaving the $(1,\frac12)\oplus
(\frac12,1)$ as the only possible chiral multiplet for these
$J=\frac32$ fields, with the transformation law
\begin{eqnarray}
\delta_5^{\vec{a}} N^\mu &=& \frac53 i \inner{a}{\tau}\gamma_5
N^\mu + 2i\gamma_5\inner{a}{\Delta_4^\mu},
\label{eq:chiralnmu}\\
\delta_5^{\vec{a}} \Delta_4^{\mu i}&=&\frac83 i\gamma_5 a^j P^{ij}_{\frac32} N^\mu
-\frac23 i \tau^i\gamma_5\inner{a}{\Delta_4^\mu}+i\inner{a}{\tau}\gamma_5
\Delta_4^{\mu i},
\label{eq:chiraldelmu}
\end{eqnarray}
where $N^\mu = N_3^\mu, \; N_4^\mu$,
and $\Delta_4^{\mu i}$ are chiral partners
belonging to the chiral multiplet $(1,\frac12)\oplus (\frac12,1)$.
Their Abelian chiral transformations are given by
\begin{eqnarray}
\delta_5 N^\mu = i \gamma_5 a N^\mu; ~~~(N^\mu = N_3^\mu,\;
 N_4^\mu),
\end{eqnarray}
hence all of these (linear combinations of) fields belong to the
single-naive Abelian representation.

Next we proceed to baryons constructed from the tensor diquark. As
mentioned earlier, the chiral transformation does not change the
spin. Due to the Fierz identity Eq.~(\ref{eq:fierzd1}),
$\Delta_5^i$ the chiral partner of the $N_5$ vanishes identically
and $N_5$ can only belong to the fundamental representation; thus
we obtain
\begin{eqnarray}
\delta_{5}^{\vec{a}}N_5 = i\inner{\tau}{a}\gamma_5 N_5.
\end{eqnarray}
For the Abelian chiral transformation
\begin{eqnarray}
\delta_5 N_5 = 3i\gamma_5 a N_5,
\end{eqnarray}
they transform as triple-naive.

Thus we have considered all possible nucleon operators:
$N_1,N_2,\cdots, N_5$.
We found two sets of two non-vanishing operators: $(N_1,N_2)$ and
$(N_{4},N_5)$; each pair belonging to the fundamental
representation $(\frac12,0)\oplus (0,\frac12)$. Hence the nucleon
operators cannot have any chiral partners. Moreover, the Fierz
identities Eqs.~(\ref{eq:fierzb1}) and (\ref{eq:fierzn1}), impose
the equivalence of the two pairs. Thus, they restrict the allowed
fields to only two, e.g. $(N_1 \pm N_2) \equiv (N_{4},N_5)$.
The Abelian chiral transformation properties distinguish between
the two remaining nucleon operators $(N_1 \pm N_2)$. Cohen and
Ji~\cite{CohenJi} have also pointed out the fact that the
$I(J)=\frac12(\frac12)$ nucleon operators belong only to the
fundamental representation. It is important to note that it is the
Pauli principle that forbids the higher dimensional chiral
representation for the nucleon operators.

For the spin 3/2 fields in the Lorentz representation
$D(\frac12,1)_{I=\frac12, \frac32}$, we obtain
\begin{eqnarray}
\delta_5^{\vec{a}}N_5^{\mu}&=&\frac53 i\inner{\tau}{a}\gamma_5
N_5^{\mu}
-4i \gamma_5\inner{a}{\Delta_5^\mu},\\
\delta_5^{\vec{a}}\Delta_5^{\mu i}&=&-\frac43 i\gamma_5 a^j
P^{ij}_{3/2} N_5^\mu -\frac23
i\tau^i\gamma_5 \inner{a}{\Delta_5^\mu}+i\inner{a}{\tau}\gamma_5 \Delta_5^{\mu i} .
\end{eqnarray}
Therefore these fields also belong to the $(1,\frac12)\oplus
(\frac12,1)$ representation. The Fierz identities
Eqs.~(\ref{eq:fierznmu2}) and (\ref{eq:fierznmu3}), however,
ensure that the two sets of $(1,\frac12)\oplus (\frac12,1)$ baryon
operators, $(N_4^\mu,\;\Delta_4^{\mu i})$ and $(N_5^\mu,\;\Delta_5^{\mu i})$, 
are actually equivalent. Their Abelian chiral
transformations are given as
\begin{eqnarray}
\delta_5 N =i\gamma_5 a N, (N=N_5^\mu,\; \Delta_5^{\mu i}),
\end{eqnarray}
hence they have the single-naive Abelian  properties. So, it turns
out that the only allowed chiral multiplet for the $D(1,\frac12)$
baryons is the $(1,\frac12)\oplus (\frac12,1)$ one, in which case
the $I=\frac32$ and $I=\frac12$ baryons are chiral partners. It
has been shown in QCD sum rules~\cite{Lee:2002jb,Lee:2006bu} that
the $\Delta_{5}^\mu$ operator gives a good description of the
$\Delta(1232)$,
but its chiral partners, the baryons with $I(J)=\frac12(\frac32),
\frac32(\frac32)$ were not discussed there. On the other hand,
lattice QCD studies~\cite{Zanotti:2003fx,Leinweber:2004it} show
that the $\Delta_4^\mu$ and $N_4^\mu$ operator describe the
$\Delta(1232)$ and the $N(1520)$ resonance.

There is is another candidate~\cite{Chung:1981cc} for the
$\Delta(1232)$ interpolating field operator: the
$D(\frac32,0)_{I=\frac32}$ representation Eq. (\ref{e:chung});
here we obtain
\begin{eqnarray}
\delta_5^{\vec{a}}\Delta_5^{\mu\nu i} = i\inner{\tau}{a}\gamma_5
\Delta_5^{\mu\nu i} - 2\epsilon^{ijk} a^j \gamma_5 \Delta_5^{\mu
\nu k} + \frac23 i\tau^i \gamma_5 \inner{a}{\Delta_5^{\mu\nu}}.
\end{eqnarray}
Therefore the $\Delta_5^{\mu\nu i}$ belongs to the
$(\frac32,0)\oplus (0,\frac32)$ chiral multiplet, and it has no
chiral partner. The Abelian chiral transformation is given by
\begin{eqnarray}
\delta_5 \Delta_5^{\mu\nu i}=3i\gamma_5 a \Delta_5^{\mu \nu i},
\end{eqnarray}
hence it is triple-naive Abelian field.

We have studied the chiral transformations of the spin-$\frac12$
and $\frac32$ baryon operators. Due to our restriction to two
light flavors, it turns out that the allowed chiral multiplets of
three-quark baryon operators have the same labels/dimensions as
the Lorentz representations that they belong to: $D(\frac12,0)$
baryons belong to the $(\frac12,0)$ chiral multiplet, the
$D(1,\frac12)$ baryons belong to the $(1,\frac12)$ and the
$D(\frac32,0)$ baryon operators belong to the $(\frac32,0)$. On
the other hand, their Abelian chiral transformation properties
depend on other properties of the operators, namely the number of
right and left components of the quark field, as explained below.

The $D(1,\frac12)_{I=\frac12,\frac32}$ chiral partners of the
$\Delta$ are of special interest. Equations (\ref{eq:chiralnmu})
and (\ref{eq:chiraldelmu}) determine, to lowest order in
perturbation theory, their (bare) axial-coupling constant matrix.
In order to see this explicitly, we consider their flavor
components, which are obtained as
\begin{eqnarray}
N_4^\mu &=&\left(\begin{array}{c} \sqrt{6}\phi_{\frac12,\frac12}\\
\sqrt{6}\phi_{\frac12,-\frac12}
\end{array}\right),\\
\Delta^{\mu +}_4 &=& \left(\begin{array}{c}
-\sqrt{\frac23} \phi_{\frac32,-\frac12}\\
-\sqrt{2}\phi_{\frac32,-\frac32} \end{array}\right),
\Delta^{\mu -}_4 = \left(\begin{array}{c}\sqrt{2}\phi_{\frac32,\frac32}\\
\sqrt{\frac23}\phi_{\frac32,\frac12}\end{array}\right),
\end{eqnarray}
where  $\Delta^{\mu \pm}_4=(\tilde{q}\gamma_\nu\gamma_5\tau^j q)
\Sprot{\mu}{\nu}\gamma_5\Iprot{\pm}{j} q$,
and the third component is eliminated by the subsidiary
condition $\tau^3\Delta^{\mu 3}_4 = -\tau^1 \Delta^{\mu 1}_4-
\tau^2\Delta^{\mu 2}_4$. $\phi_{I,I_Z}$ are the properly normalized flavor
wave-functions with the given isospin. In terms of the axial
rotation about the third axis $\vec{a}=(0,0,a_3)$,  we obtain the
axial-coupling constant matrix from Eqs.~(\ref{eq:chiralnmu}) and
(\ref{eq:chiraldelmu}),
 e.g. for $\phi_{\frac12,\frac12}$, $\phi_{\frac32,\frac12}$ and
$\phi_{\frac32,\frac32}$,
\begin{eqnarray}
\delta_5^{a_3} \phi_{\frac32,\frac32}&=&i\gamma_5 a_3
\phi_{\frac32,\frac32},\\
\delta_5^{a_3}\left(\begin{array}{c} \phi_{\frac12,\frac12}\\
\phi_{\frac32,\frac12}\end{array}\right)
&=&i\gamma_5 a_3 \left(\begin{array}{cc} \frac53 & -\frac{2\sqrt{2}}{3}\\
-\frac{8\sqrt{2}}{3} & \frac13 \end{array}\right)
\left(\begin{array}{c} \phi_{\frac12,\frac12}\\
\phi_{\frac32,\frac12}\end{array}\right),
\end{eqnarray}
with the familiar (``$SU_{\rm FS}(6)$") value $\frac53$ for its
``nucleon" component. Manifestly, this value was not obtained
using the $SU_{\rm FS}(6)$ symmetry, but from the chiral $SU(2)_R
\times SU(2)_L$ symmetry.

\section{Summary and Conclusions}
We have investigated the chiral multiplets consisting of local
three-quark baryon operators, where we took into account the Pauli
principle by way of the Fierz transformation. All spin 1/2 and 3/2
baryon operators were classified according to their Lorentz and
isospin group representations, where spin and isospin projection
operators were employed.

We derived all non-trivial relations between various baryon
operators due to the Fierz transformations, and thus found the
independent baryon fields. We showed that the Fierz transformation
connects only the baryon operators with identical
group-theoretical properties, i.e., belonging to the same chiral
multiplet. Then we studied chiral transformations of the baryon
operators, where the Fierz identities restrict the permissible
baryons' chiral multiplets.

We also found that baryons with different isospins may mix under
the chiral transformations, i.e., they may belong to the same
chiral multiplet, whereas baryons with different spins can not.
The parity does not play an apparent role in the chiral properties
of the baryon operators. Thus, naively, the nucleon could have a
chiral partner - a baryon operator with
$I(J^{P})=\frac32(\frac12^{\pm})$. This chiral partner vanishes
identically, however, due to the Pauli principle/Fierz identities.

There are two possible choices for the baryon operator with $J =
3/2$; one is $D(1, \frac12)$ and the other is $D(\frac32, 0)$. In
the former choice, the baryon isodoublet with
$I(J)=\frac12(\frac32),\; \frac32(\frac32)$ are the $\Delta$'s
chiral partners. In the latter choice, the $\Delta$ does not have
a chiral partner. The two $\Delta$ fields also differ in their
Abelian chiral properties: the $(1,\frac12)\oplus (\frac12,1)$
multiplet has Abelian axial charge -1 and the $(\frac32,0) \oplus
(0,\frac32)$ multiplet has Abelian axial charge +3. The Abelian
($U_A(1)$) chiral properties also resolve the nucleon field
ambiguity (Ioffe \cite{Ioffe81}): one nucleon has Abelian axial
charge +3; another has Abelian axial charge -1. Physical
consequences of this $U_A(1)$ ambiguity have to some extent
already been explored in Ref. \cite{NHD1} in the case of nucleons,
whereas the question remains completely unexplored in the case of
the $\Delta$.

Some partial results have already been pointed out by Ioffe
\cite{Ioffe81}, by Chung, Dosch, Kremmer and Schall
\cite{Chung:1981cc}, and by Cohen and Ji \cite{CohenJi}, but they
were limited in their scope. For example, Ioffe \cite{Ioffe81} and
Chung, Dosch, Kremmer and Schall \cite{Chung:1981cc}, were
concerned only with the Fierz relations and the number of
independent fields, not with their chiral properties. Moreover,
Ioffe did not mention the $D(\frac32,0)_{I=\frac32}$ spin 3/2
fields - he only considered the Lorentz representation
$D(\frac12,1)_{I=\frac32}$ spin 3/2 fields; only Chung et al.
\cite{Chung:1981cc} considered the $D(\frac32,0)_{I=\frac32}$ spin
3/2 fields. Cohen and Ji \cite{CohenJi}, on the other hand, were
concerned primarily with the chiral properties. We have unified
these two questions and showed that only a complete analysis leads
to a meaningful answer to both. Our results answer a specific
question raised recently in the study of baryon chiral multiplets,
\cite{Beane:2002td}, and thus open the door to further explicit
studies in this field.

We have employed the standard isospin formalism instead of the
explicit expressions in terms of different flavored quarks in the
flavor components of the baryon fields that are commonplace in
this line of work. By using the isospin formalism, we have been
able to derive all Fierz identities and chiral transformations of
the baryons systematically. The extension to SU(3) is not as
straightforward as one might have imagined, however, so we leave
it for another occasion.

\section*{Acknowledgments}
\label{ack}

We wish to thank Dr. N. Ishii and Prof. W. Bentz, for valuable
conversations about the Fierz transformation of triquark fields
and (in)dependence of the nucleon interpolating fields. We also
thank Prof. D. Jido, for the fruitful discussions on the Fierz
transformation in the $L-R$ representation. One of us (V.D.)
wishes to thank Prof. S. Fajfer for hospitality at the Institute
Jo\v zef Stefan, where this work was started and to a large extent
completed, and to Prof. H. Toki for hospitality at RCNP, where it
was finalized.

\appendix
\section{Fierz Transformation}
\label{app}
We summarize the detailed results of the Fierz transformation of
baryons.
After the Fierz transformation of the iso-spin, Dirac and color, 
the Fierz transformed field ${\cal F}[N]$ satisfy the relation 
$N=-{\cal F}[N]$, namely the Pauli principle. 

For $D(\frac12,0)$ and $I=\frac12$,
\begin{eqnarray}
{\cal F}\left[\left(\begin{array}{c}
N_1 \\ N_2 \\ N_3\\ N_4 \\ N_5 \end{array}\right)\right]=-
\frac18\left(\begin{array}{ccccc}
1   & 1  & 1  & -1 & \frac12\\
1   & 1  & -1 & 1  & \frac12\\
4   & -4 & -2 & -2 & 0 \\
-12 & 12 & -6 & 2  & 0 \\
36  & 36 & 0  & 0  & 2
\end{array}
\right)\left(\begin{array}{c}
N_1 \\ N_2 \\ N_3\\ N_4 \\ N_5 \end{array}\right)_.
\end{eqnarray}

For the $D(\frac12,0)$ and $I=\frac32$ operators,
\begin{eqnarray}
{\cal F}[\Delta_4^i]&=&\frac12 \Delta_4^i,\\
{\cal F}[\Delta_5^i]&=&-\frac12 \Delta_5^i.
\end{eqnarray}

For the $D(\frac12,1)$ and $I=\frac12$ operators,
\begin{eqnarray}
{\cal F}\left[\left(\begin{array}{c}
N_{3\mu} \\ N_{4\mu} \\ N_{5\mu}\end{array} \right)\right]
=-\frac18 \left(\begin{array}{ccc} 2 & 2 & -2 \\
6 & -2 & -2 \\ -12 & -4 & 0 \end{array}\right)
\left(\begin{array}{c}
N_{3\mu} \\ N_{4\mu} \\ N_{5\mu}\end{array} \right)_.
\end{eqnarray}

For the $D(\frac12,1)$ and $I=\frac32$ operators,
\begin{eqnarray}
{\cal F}\left[\left(\begin{array}{c} \Delta^{i\mu}_4 \\
\Delta^{i\mu}_5 \end{array}\right)\right]
=-\frac14 \left( \begin{array}{cc} 2 & 2 \\ 4 & 0 \end{array}\right)
\left(\begin{array}{c} \Delta^{i\mu}_4 \\ \Delta^{i\mu}_5 \end{array}\right)_.
\end{eqnarray}

For the $D(\frac32,0)$ operators,
\begin{eqnarray}
{\cal F}\left[N_5^{\mu\nu}\right]&=&\frac12 N_5^{\mu\nu},\\
{\cal F}\left[\Delta_5^{\mu\nu i}\right]&=&-\Delta_5^{\mu\nu i}.
\end{eqnarray}

\baselineskip 5mm
\bibliographystyle{h-physrev4}

\end{document}